\documentclass[12pt]{article}
\usepackage[dvips]{graphicx}
\title{ Renormalization Analysis of Resonance Structure in 2-D
Symplectic Map
   }
\author{Tsuyoshi 
Maruo$^a$\footnote{e-mail:~tmaruo@allegro.phys.nagoya-u.ac.jp},\quad 
Shin-itiro
Goto$^b$\footnote{e-mail:~sgoto@amp.i.kyoto-u.ac.jp},
\quad Kazuhiro Nozaki$^a$ \\
{\small $a$ Department of Physics, }\\
{\small Nagoya University, Nagoya 464-8602, Japan}\\
{\small $b$ Department of Applied Mathemaics and Physics,}\\
{\small Kyoto University, Kyoto, 606-8501, Japan}
}
\date{}
\makeindex

\begin{document}
\maketitle

\newcommand{\beq}{\begin{equation}}
\newcommand{\beqa}{\begin{eqnarray}}
\newcommand{\eeq}{\end{equation}}
\newcommand{\eeqa}{\end{eqnarray}}
\newcommand{\non}{\nonumber}
\newcommand{\lb}{\label}
\newcommand{\fr}[1]{(\ref{#1})}
\newcommand{\cc}{\mbox{c.c.}}
\newcommand{\nr}{\mbox{n.r.}}
\newcommand{\tx}{\widetilde{x}}
\newcommand{\tg}{\widetilde{g}}
\newcommand{\hx}{\widehat{x}}
\newcommand{\tA}{\widetilde A}
\newcommand{\tB}{\widetilde B}
\newcommand{\tK}{\widetilde K}
\newcommand{\tc}{\widetilde c}
\newcommand{\tAc}{{\widetilde A}^{*}}
\newcommand{\tphi}{{\widetilde \phi}}
\newcommand{\btA}{\mbox{\boldmath {$\widetilde A$}}}
\newcommand{\bA}{\mbox{\boldmath {$A$}}}
\newcommand{\bC}{\mbox{\boldmath {$C$}}}
\newcommand{\bu}{\mbox{\boldmath {$u$}}}
\newcommand{\bN}{\mbox{\boldmath {$N$}}}
\newcommand{\bZ}{\mbox{\boldmath {$Z$}}}
\newcommand{\bR}{\mbox{\boldmath {$R$}}}
\newcommand{\ve}{{\varepsilon}}
\newcommand{\e}{\mbox{e}}
\newcommand{\Aba}{\bar{A}}
\newcommand{\aCos}{\mbox{Cos}^{-1}}
\newcommand{\aTan}{\mbox{Tan}^{-1}}
\newcommand{\mapright}[1]{%
    \smash{\mathop{%
       \hbox to 2cm{\rightarrowfill}}\limits^{#1}}}
\newcommand{\mapleft}[1]{%
    \smash{\mathop{%
       \hbox to 2cm{\leftarrowfill}}\limits^{#1}}}
\def\lmapdown#1{\Big\downarrow
    \llap{$\vcenter{\hbox{$\scriptstyle#1\,$}}$ }}
\def\rmapdown#1{\Big\downarrow
    \rlap{$\vcenter{\hbox{$\scriptstyle#1\,$}}$ }}

%%%%%%%%%%%%%%%%
\begin{abstract}
%%%%%%%%%%%%%%%%%
A symplecticity-preserving RG analysis is carried out  to study a
resonance structure near an elliptic fixed point of a proto-type
symplectic map in two dimensions.  Through analyzing fixed points
of a reduced RG map, a topology of the resonance structure such as a
chain of resonant islands
can be determined analytically. An application of this analysis to the
H\'enon map is also presented.\\
PACS-1995:
03.20.+i ,
%(classical mechanics of discrete systems: general mathematical aspects)
47.20.Ky,
%(Nonlinearity (including bifurcation theory)
02.30.Mv,
%(Approximations and expansions)
64.60.Ak.
%(Renormalization-group, fractal, and percolation studies of phase transitions)

Keywords: symplectic mappings, renormalization group method, H\'enon map,
resonant islands
%%%%%%%%%%%%%%
\end{abstract}
%%%%%%%%%%%%%%

%%%%%%%%%%%
\pagebreak
%%%%%%%%%%%

%%%%%%%%%%%%%%%%%%%%%%
\section{Introduction}
%%%%%%%%%%%%%%%%%%%%%%
Since the renormalization group (RG) method was developed as an
asymtotic method to study a long-time behaviour of a flow in a
dynamical system, \cite{CGO96} \ there have been several attempts to
extend the method to apply to  discrete systems. \cite{KM98,GN01Prog}
Particularly, the application to  a symplectic map is of great
importance since
a symplectic map is generally derived as a Poincare map of a
Hamiltonian flow describing
a physical system. However, a naive application of the RG method  fails
to describe a long-time behaviour of a  system since the naive
renormalization process does not secure the symplectic smmmetry to an
RG map  \cite{GN01JPSJ} . This difficulty generally arises in a
discrete system but does not occur in the application to a Hamiltonian
flow.  Therefore, it is urgent to develop the RG method
for a discrete system which preseves the symplectic symmetry.  After
some attempts to treat
special cases \cite{GN01JPSJ},\cite{GNY02}, a general procedure
has  been proposed recently in order to  derive a
symplecticity-preserving (symplectic) RG map
\cite{GN03}. The newly proposed procedure is based on the observation
that  a naive RG map
is not symplectic but a finite truncation of the so-called Loiville
expansion of a Hamiltonian flow.
Then, a symplectic RG map is constructed by means of a symplectic
integrator of the underlying Hamiltonian flow.

Employing  the newly proposed simplecticity-preserving procedure in the
RG method, we analyze  resonance structures such as a chain of resonant
islands near an elliptic fixed point in a two-dimensional symplectic
map.  A chain of resonant islands is characterized by alternating
positioning of  hyperbolic and elliptic periodic points in the phase
space of a symplectic map. In our
analysis, these periodic points are shown to be correspond to the same
number of fixed points of
an RG map, which simplifies the analysis to prove existence of a chain
of resonant islands.
As an application of the present analysis, resonance structures of the
H\'enon map \cite{henon} are reproduced by means of symplectic RG maps. A
similar analysis to the H\'enon map was tried using the RG method
\cite{TD03}. However, the previous study did not succeed in
    reconstruction of  resonance structures of the H\'enon map in terms of
a symplectic RG map, which would be
the most impressive result of an application of the RG method.
Here, as an example of  general results , we carry out a comprehensive
study of resonance structures of the H\'enon map within the framework of
the RG method.

%%%%%%%%%%%%%%%%%%%%%%%%%%%%%%%%%%%%%%%%%%%%%%%%%%%%%%
\section{Symplectic Map near Elliptic Fixed point}
%%%%%%%%%%%%%%%%%%%%%%%%%%%%%%%%%%%%%%%%%%%%%%%%%%%%%%%
Let us start with the following symplectic map of action-angle type
, $(x_n,y_n)\mapsto (x_{n+1},y_{n+1})$:
\beqa
x_{n+1}&=&x_n+y_{n+1}\non\\
y_{n+1}&=&y_n+f(x_n).\lb{o-sympmap}
\eeqa
The origin $(x_n,y_n)=(0,0)$ is assumed to be an elliptic fixed point
of the map (\ref{o-sympmap}).
Expanding $f(x_n)$ around the fixed point, we have
\beqa
x_{n+1}&=&x_n+y_{n+1}\non\\
y_{n+1}&=&y_n+\sum_{m=1}a_m x_n^m ,\lb{o-sympmap1}
\eeqa
where $a_m$ are constants. Since the origin  is supposed to be elliptic,
there are two eigenvalues at the origin: $\exp (\pm i \omega)$, where
$\omega$ is real
and given in terms of the coefficient  $a_1$  as
\beq
2\cos \omega=2+a_1,
\eeq
where $0\ge a_1\ge -4$.
Eliminating $y_n$ from  Eq.(\ref{o-sympmap1}) , we get
\beq
x_{n+1}-2\cos \omega x_n +x_{n-1}=a_2 x_n^2+a_3x_n^3+\cdots.
\lb{basicmap}
\eeq
In the case  $a_m=0$ for $m>2$, the map (\ref{basicmap}) is called the
H\'enon map  \cite{henon}.
If only the coefficient $a_3$ is retained while all other nonlinear
coefficients vanish,   Eq.(\ref{basicmap} ) becomes the double-well
type map studied in \cite{GN01JPSJ}, \cite{GN03}.
The resonance structure near an elliptic fixed point is studied by
taking Eq.(\ref{basicmap}) as a
prototype in the following sections.

%%%%%%%%%%%%%%%%%%%%%%%%%%%%%%%%%%%%%%%%%%%%%%%%%%%%%
\section{Perturbative Rrenormalization Analysis of Resonance Structure}
%%%%%%%%%%%%%%%%%%%%%%%%%%%%%%%%%%%%%%%%%%%%%%%%%%%%%%%

Resonance structures in the phase plane of the prototype map
(\ref{basicmap})are analyzed by the
perturbative RG method.  We focus our attention to the case where the
frequency $\omega$ is close to
one of  resonant frequencies $2\pi/k, k=3,4,5$:
\beq
\omega=2\pi/k+\ve^p\delta, \lb{resonace-detune}
\eeq
in which $\ve$ ia a small parameter representing small resonance
detuning  and $p(>0)$ will be chosen later so that the small resonamce
detuning balances with  a nonlinear term of the leading order.
Then, the map (\ref{basicmap}) reads
\beq
x_{n+1}-2\cos (2\pi/k) x_n +x_{n-1}=2\{\cos \omega-\cos (2\pi/k)
\}x_n+a_2 x_n^2+
a_3x_n^3+\cdots, \lb{basicmap1}
\eeq
where the first term of the RHS of Eq.(\ref{basicmap1}) represents
small resonance detuning. %\\

Near the origin of the phase plane, assume $x_n$ be expanded as
\beq
x_n=\ve  x_n^{(1)}+\ve^2  x_n^{(2)}+\ve^3
x_n^{(3)}+\cdots,\lb{expansion}
\eeq
then the first order perturbation equation gives
\beq
L x_n^{(1)}\equiv x_{n+1}^{(1)}-2 x_n^{(1)} \cos (2\pi/k)
+x_{n-1}^{(1)}=0,
\eeq
from which we have
\beq
x_n^{(1)}=A\exp\{i(2\pi/k) n\}+\Aba\exp\{-i(2\pi/k) n\}, \lb{first-sol}
\eeq
where $A$ is a complex constant and $\Aba$ is the complex conjugate to
$A$.
The ordinary perturbation analysis provides
$ x_n^{(m)} \quad (m\ge 2)$ in terms of polynomials of $A$ and $\Aba$ .
Among various polynomials, secular polynomials,  of which coefficients
depend
on some powers of $n$,  play a crucial role for a long-time behaviour
of the system.
In our RG method, a special transformation for $A$ called an RG
transformation  is introduced
\cite{GMN99},\cite{GN01JPSJ},\cite{rgtrans}
   and an RG map is constructed for the
renormalized $A$ so that such secularity is removed.
Let us estimate the order of magnitude of such secular polynomials. We
generally encounter
the following type of secular polynomial  known as  a term of nonlinear
frequency shift.
\beq
(C_1 n |A|^2+(C_2n+C_3n^2)|A|^4+\cdots)A\exp(i\omega n), \lb{non-freq}
\eeq
where $C_m  ( m=1,2,3)$ are constants.  Thus, the term of nonlinear
frequency shift
is of the order $\ve^3$ or higher.  %\\

Since the frequency  of the leading order solution (\ref{first-sol}) is
   the
   resonant frequency $2\pi/k$, the following secular term appears, in
addition to the nonlinear
   frequency shift term (\ref{non-freq}),
   \beq
   C_4 n \Aba^{(k-1)}, \lb{resonant}
   \eeq
   where $C_4$ is a constant and $k$ corresponds to the resonant
frequency $2\pi/k$.
   In this paper, we call the secular term (\ref{resonant}) a resonant
secular term.
   The order of the resonant secular term is $\ve^{(k-1)}$ .
Since  $\Aba$ is the coefficient of the first order solution
$\Aba\exp\{-i(2\pi/k) n\}$ and
\beqa
[\Aba\exp\{-i(2\pi/k) n\}]^{(k-1)}&=&\Aba^{(k-1)}\exp\{-i(2\pi(k-1)/k)
n\} \non \\
&=&\Aba^{(k-1)}\exp\{i(2\pi/k) n\}, \non
\eeqa
    the $(k-1)$ -th order nonlinear term $[\Aba\exp\{-i(2\pi/k)
n\}]^{(k-1)}$
causes a secular solution.
For $k> 4$,  the term of nonlinear frequency shift  (\ref{non-freq})
is a dominant nonlinear term while the resonant secular term
(\ref{resonant})  yields small  but important contribution .
Since  the resonance detuning $\ve^p\delta $ should  balance with the
dominant
nonlinear term,  $p$ should be chosen as $p=2$ in this case and for
$k=4$.
In the case  $k= 3$ , the resonant secular term dominates over the the
term of nonlinear frequency shift and $p$ should be chosen as $p=1$ so
that the resonant secular term balances with
the resonance detuning term. \\
In the following subsections, we carry out the  perturbative RG
analysis of (\ref{basicmap1}) for
   resonant frequencies $2\pi/k (k=3,4,5)$.
%%%%%%%%%%%%%%%%%%%%%%%%%%%
\subsection{Resonant Frequency : $2\pi/3$}
%%%%%%%%%%%%%%%%%%%%%%%%%%
For $k=3$, the resonant frequency becomes $2\pi/3$ and  the small
resonance detuning is
chosen to be of the order $\ve$ as
$$
\omega-\omega _0=\ve\delta,
$$
where $\omega _0=2\pi/3$.
Substituting the expansion (\ref{expansion}), where the first order
solution $ x_n^{(1)}$ is given by
Eq.(\ref{first-sol}),  into the map (\ref{basicmap1}), we have the
following second order
equation.
\beq
L x_n^{(2)}=-2\delta\sin \omega _0 \{A\exp(i\omega _0 n)+\cc\}
+a_2\{A^2\exp(2i\omega _0 n)
+\cc+2|A|^2\},\lb{second-eq}
\eeq
where   $\cc$ stands for the complex conjugate to the preceding term(s).
Since  $\exp(\pm 2i\omega _0)=\exp(\mp i\omega _0)$ for  $\omega
_0=2\pi/3$,
the nonlinear terms  proportional to $A^2$ and $\Aba^2$ and the linear
detuning term
in Eq.(\ref{second-eq}) causes secular terms in the second order
solution $x_n^{(2)}$ :
\beq
x_n^{(2)}=i \bigl( \delta A-\frac{a_2}{2\sin \omega _0} \Aba^2 \bigr) n
\exp(i\omega _0 n)
+\cc  +a_{2,0} |A|^2, \lb{second-sol}
\eeq
where
\beq
a_{2,0}=\frac{a_2}{1-\cos \omega _0}. \lb{a_20}
\eeq
Up to the second order in $\ve$, we have
\beq
x_n/\ve =\Bigl( A+ i \ve\bigl( \delta A-\alpha \Aba^2 \bigr) n \Bigr)
\exp(i\omega _0 n)   +\cc + \ve  a_{2,0} |A|^2,\lb{up-second-sol}
\eeq
where
$$\alpha =\frac{a_2}{2\sin \omega _0}.$$
In order to remove secularity in Eq.(\ref{up-second-sol}) and derive a
symplecticity-preserving
RG map,  we follow the procedure established in \cite{GN03} or
\cite{GNY02}.
First, we introduce the
renormalization transformation $A\mapsto A_n$ :
\beq
A_n= A+ i \ve\bigl( \delta A-\frac{a_2}{2\sin \omega _0} \Aba^2 \bigr)
n,
\eeq
from which we obtain a naive RG map $A_n\mapsto A_{n+1}$.
\beq
A_{n+1}=A_n+i\ve ( \delta A_n-\alpha  \Aba_n^2 )
+{\cal O}(\ve^2). \lb{naive-rg1}
\eeq
Then, from Eq.(\ref{up-second-sol}), the original variable $x_n$ is
rewritten in terms of the renormalized amplitude $A_n$ as
\beq
x_n/\ve =A_n \exp(i\omega _0 n)   +\cc + \ve  a_{2,0} |A_n|^2.
\lb{up-second-sol2}
\eeq
The naive RG map (\ref{naive-rg1}) is not symplectic but is made
symplectic by means of the symplectic integration method.
It should be noted that the map (\ref{naive-rg1}) is the first order
truncation
of the following  Liouville  expansion of a Hamiltonian flow
\beq
A(t+\ve)=\exp(\ve{\cal L}_H)A(t)=\bigg(1+\ve{\cal
L}_H+\frac{\ve^2}{2!}{\cal L}^2_H+\cdots \bigg)A(t),
\eeq
with a Hamiltonian
\beq
H( A(t),\Aba(t))=i\bigl(\delta A(t)\Aba(t)-\frac{a_2}{6\sin\omega
_0}(A(t)^3+\Aba(t)^3)\bigr). \lb{hamil3}
\eeq
where $A(t+\ve)=A_{n+1}, A(t)=A_n$ and the Liouville operator ${\cal
L}_H$  is defined, in terms of the Poisson bracket $\{ , \}$,  as
\beqa
{\cal L}_H A&\equiv&\{A,H\},\non \\
{\cal L}_H^2 A&=&{\cal L}_H\bigg({\cal L}_H A\bigg)=\{\{A,H\},H\}.\non
\eeqa
Therefore, the Hamiltonian ( \ref{hamil3}) describes the phase space
structure of the renormalized amplitude $(A_n, \Aba_n)$  within the
order of the present approximation, regardless of  choise of
symplectic integrators  to the  Hamiltonian $H$.  The most important
points in the the phase space  $(A_n, \Aba_n)$ are fixed points since
they ,via Eq.(\ref{up-second-sol2}), correspond to periodic points in
the original phase space
$(x_n,x_{n+1})$, which  determine  resonance structure such as a chain
of resonant islands .
The fixed points to the flow of $H$ are solutions of the following
algebraic equations
\beq
\frac{\partial H(A_n,\Aba_n)}{\partial A_n}=0, \qquad
\frac{\partial H(A_n,\Aba_n)}{\partial \Aba_n}=0.
\lb{fixed-eq}
\eeq
In the present case, Eqs.(\ref{fixed-eq}) are reduced to
\beq
\delta-\frac{a_2}{2\sin\omega _0}r\exp(-3i\theta)=0, \lb{fix-eq1}
\eeq
where $A_n=r\exp(i\theta)  (r> 0)$  and we obtain three hyperbolic
fixed points
\beq
\theta=0, \frac {2\pi}{3}, \frac {4\pi}{3},  \qquad r=\frac
{2\delta\sin\omega _0 }{a_2}
\eeq
for $\delta/a_2>0$ or
\beq
\theta=\frac {\pi}{3}, \frac {3\pi}{3},\frac {5\pi}{3},
\qquad r=\frac {-2\delta\sin\omega _0 }{a_2}
\eeq
for $\delta/a_2<0$. These fixed points are also those of the naive RG
map (\ref{naive-rg1})
and symplectic RG maps derived later.
Thus, we find that the resonance structure in the
original phase plane  is characterized by the three hyperbolic periodic
points with a period 3
instead of a chain of resonant islands.    Higher order corrections to
the dominant Hamiltonian
( \ref{hamil3}) slightly modify the values of fixed points and do not
change this result as a whole. Two symplectic RG maps are
   derived from the naive RG map (\ref{naive-rg1}) by choosing  
approriate
canonical variables to the Hamiltonian. For canonical varibles  $(A'_n,
A''_n )$ where
   $A_n=A'_n+i A''_n, \quad  A'_n, A''_n  \in \mbox{R}$, we have
\beqa
A'_{n+1}&=&A'_n+\ve f_3(  A'_{n+1},A''_n)
,\non \\
A''_{n+1}&=&A''_n+\ve g_3( A'_{n+1},A''_n) ,
\lb{symp-rg3}
\eeqa
where  $f_3(A',A'')=-( \delta A''+2\alpha A'A'') $ and
$g_3(A',A'')=\delta A'-\alpha( (A')^2-(A'')^2) $.
    Introducing the action-angle
variables $(J_n, \theta_n)$ where $A_n=\sqrt{J_n}\exp{i\theta_n}$, we
have another
symplectic RG map.
\beqa
J_{n+1}&=&J_n+\ve F_3(J_{n+1},\theta_n) ,\non \\
\theta _{n+1}&=&\theta _n+\ve G_3(J_{n+1},\theta_n) ,\lb{sy-rg3}
\eeqa
where $F_3(J,\theta)=-2\alpha J^{3/2}\sin(3\theta)$ and
$G_3(J,\theta)=\delta-
\alpha \sqrt{J}\cos(3\theta)$.
The former symplectic RG map (\ref{symp-rg3}) takes a set of
   explicit diference equations, while the latter map  (\ref{sy-rg3})
 exactly preserves a symmetry of $2\pi/3$ rotation which
the Hamiltonian
(\ref{hamil3}) and the naive RG map (\ref{naive-rg1}) possess.
The RG map  (\ref{sy-rg3}) in the angle-action variables was first 
derived in
\cite{TD03}.

%%%%%%%%%%%%%%%%%%%%%%%%%%%
\subsection{Resonant Frequency : $2\pi/4$}
%%%%%%%%%%%%%%%%%%%%%%%%%%
For $k=4$, the resonant frequency $\omega _0$ becomes $\pi/2$ and  the
small resonance detuning is
chosen to be of the order $\ve^2$ as
\beq
\omega-\omega _0=\ve^2\delta. \lb{detune}
\eeq
Substituting the expansion (\ref{expansion})  into the map
(\ref{basicmap1})
and taking Eq.(\ref{detune}) into account, we have a similar expression
to the  second order equation
as Eq.(\ref{second-eq}), in which the frequency detuning term shoud be
moved to the
third order equation.  Then, in contrast to the previous case, this
second order equation does not
include secular terms and we have the second order solution as a
non-secular solution
\beq
x_n^{(2)}=a_{2,2}A^2\exp(2i\omega _0 n)+\cc  +a_{2,0}|A|^2,
\lb{second-sol-4}
\eeq
where
\beq
a_{2,2}=\frac{a_2}{2(\cos(2\omega _0)-\cos\omega _0)}. \lb{a_22}
\eeq
The third order equation is written as
\beqa
L x_n^{(3)}&=&\Bigl(-2\delta\sin \omega _0
+\bigl(2a_2(a_{2,2}+a_{2,0})+3a_3\bigr)|A|^2\Bigr)A\exp(i\omega
_0 n)  \non \\
& &+(2a_2a_{2,2}+a_3)A^3\exp(3i\omega _0 n)+\cc.\lb{third-eq}
\eeqa
Since $\exp(\pm 3i\omega _0)=\exp(\mp i\omega _0)$ for $\omega
_0=\pi/2$,
all  forcing terms in Eq.(\ref{third-eq}) contribute to a secular
solution.
The third order secular solution is given as
\beq
   x_n^{(3)}=i\bigl(
(\delta+a_{3,1}|A|^2)A+a_{3,-3}\Aba^3\bigr)n\exp(i\omega _0 n)
+\cc, \lb{third-sol}
\eeq
where
\beqa
a_{3,1}&=&-\frac{1}{2\sin\omega
_0}\bigl(2a_2(a_{2,2}+a_{2,0})+3a_3\bigr), \non \\
a_{3,-3}&=&-\frac{1}{2\sin\omega _0}(2a_2a_{2,2}+a_3). \non
\eeqa
The secularity in $x_n^{(3)}$ is removed by the following
renormalization
transformation:
\beq
A_n=A+i\ve^2\bigl( (\delta+a_{3,1}|A|^2)A+a_{3,-3}\Aba^3\Bigr)n,
\eeq
which yields a naive RG map.
\beq
A_{n+1}=A_n+i\ve^2\bigl(
(\delta+a_{3,1}|A_n|^2)A_n+a_{3,-3}\Aba_n^3\Bigr). \lb{naive-rg2}
\eeq
The original variable $x_n$ is represented in terms of the renormalized
amplitude $A_n$ as
\beq
x_n/\ve=A_n \exp(i\omega _0 n)  +
\ve a_{2,2}A_n^2\exp(2i\omega _0 n)+\cc  +\ve a_{2,0}|A_n|^2.
\lb{up-third-sol}
\eeq
The naive RG map (\ref{naive-rg2}) is derived from the Liouville
expansion with a Hamiltonian
\beq
H=i\bigl(\delta |A_n|^2+\frac{a_{3,1}}{2}
|A_n|^4+\frac{a_{3,-3}}{4}(A_n^4+\Aba_n^4)\bigr).
\lb{hamil-2}
\eeq
Fixed points in this Hamiltonian flow, which are also fixed points of
the naive RG map (\ref{naive-rg2})
, are determined by the similar algebraic equations as those in the
previous subsection.
The fixed points $A_n=r\exp(i\theta)$ satisfy
\beqa
\sin(4\theta)&=&0, \\
r^2&=& -\frac{\delta}{a_{3,1}+a_{3,-3}\cos(4\theta)}>0,
\eeqa
which give eight fixed points
\beqa
r^2&=&\frac{\delta}{2a_3}, \qquad \theta =0, \frac{\pi}{2},\pi,
\frac{3\pi}{2}, \lb{fix-21}\\
r^2&=&\frac{\delta}{a_2^2+a_3}, \qquad \theta
=\frac{\pi}{4},\frac{3\pi}{4},
\frac{5\pi}{4},\pi, \frac{7\pi}{4},\lb{fix-22}
\eeqa
where the following conditions are assumed
\beq
\frac{\delta}{2a_3}>0, \qquad \frac{\delta}{a_2^2+a_3}>0. \lb{fix-cod1}
\eeq
Four fixed points (\ref{fix-21}) are found to be elliptic, while the
other four fixed points
   (\ref{fix-22})  are hyperbolic. Therefore,  if the conditions
(\ref{fix-cod1}) are satisfied,
resonance structure in this case consists of a chain of four resonant
islands.
In \cite{GN03},  such a chain of resonant islands is shown for the case
$a_2=0$.
Following the same  procedure as in the previous subsection, we derive
two symplectic RG maps.
   \beqa
A'_{n+1}&=&A'_n+\ve^2 f_4(  A'_{n+1},A''_n)
,\non \\
A''_{n+1}&=&A''_n+\ve^2 g_4( A'_{n+1},A''_n) ,
\lb{symp-rg4}
\eeqa
where  $f_4(A',A'')=-\bigl( \delta A''+(a_{3,1}+a_{3,-3})(A'')^3
+(a_{3,1}-3a_{3,-3})(A')^2A''\bigr) $ and
$g_4(A',A'')=\delta A'+(a_{3,1}+a_{3,-3})(A')^3
+(a_{3,1}-3a_{3,-3})(A'')^2A' $.
This symplectic RG map takes a set of explicit difference eqations in
the case $a_2=0$
\cite{GN03}. Another one  preserves a symmery of $2\pi /4$ rotation
exactly.
\beqa
J_{n+1}&=&J_n+\ve^2 F_4(J_{n+1},\theta_n) ,\non \\
\theta _{n+1}&=&\theta _n+\ve^2 G_4(J_{n+1},\theta_n) ,
\lb{sy-rg4}
\eeqa
where $F_4(J,\theta)=2a_{3,-3} J^2\sin(4\theta)$ and
$G_4(J,\theta)=\delta+a_{3,1}J
+a_{3,-3}J\cos(4\theta)$.

%%%%%%%%%%%%%%%%%%%%%%%%%%%
\subsection{Resonant Frequency : $2\pi/5$}
%%%%%%%%%%%%%%%%%%%%%%%%%%
When the resonant frequency $\omega _0$ is $2\pi/5$,  the small
resonance detuning is
chosen to be of the order $\ve^2$. Then, the similar analysis as in
previous sections yields
the following perturbation solution.
While $x_n^{(2)}$ is given by Eq.(\ref{second-sol-4}), $x_n^{(3)}$  is
obtained as
\beq
   x_n^{(3)}=i\big((\delta+a_{3,1}|A|^2)A n \exp(i\omega _0 n)\bigr)
   +a_{3,3}A^3\exp(3i\omega _0 n)+ \cc, \lb{third-sol}
   \eeq
   where
   $$a_{3,3}=\frac{2a_2a_{2,2}+a_3}{2(\cos(3\omega _0)-\cos\omega _0)}$$
   and  the secular part of $x_n^{(4)}$ is determined by
   \beqa
Lx_n^{(4)}&=&\bigl((2a_2a_{3,3}+a_2a_{2,2}^2+3a_3a_{2,2}+a_4)\Aba^4\bigr
)
   \exp(i\omega _0 n) \non \\
   & &+2ia_2(\delta+a_{3,1}|A|^2)A^2 n \exp(2i\omega _0 n)+\cc +
\mbox{non-secular terms},
   \lb{fourth}
    \eeqa
where a relation $ \exp(\pm 4i\omega _0 )=\exp(\mp i\omega _0 )$ has
been  used.
A secular solution of Eq.(\ref{fourth}) yields the secular part of
$x_n^{(4)}$:
\beq
x_n^{(4)}=-i\alpha _4 \Aba^4 n \exp(i\omega _0 n)+ia_{4,2}(\delta
+a_{3,1}|A|^2)A^2
n\exp(2i\omega _0 n),\lb{forth-sol}
\eeq
where
\beqa
\alpha _4&=&\frac{2a_2a_{3,3}+a_2a_{2,2}^2+3a_3a_{2,2}+a_4}{2\sin\omega
_0},\non \\
a_{4,2}&=&\frac{a_2}{\cos(2\omega _0)-\cos\omega _0}. \non
\eeqa
All secular terms in Eqs. (\ref{third-sol}) and (\ref{forth-sol}) are
removed by introducing
a renormalization transformation.
\beq
A_n=A+i\ve^2(\delta+a_{3,1}|A|^2)A n-i\ve^3\alpha_4 \Aba^4 n.
\lb{renom-5}
\eeq
Then, we finally obtain
a secular-free perturbation solution of the original equation
(\ref{basicmap1})  up to
${\cal O} (\ve^3)$:
\beqa
x_n&=&\ve (A_n \exp(i\omega _0 n)+ 
\cc)+\ve^2\bigl((a_{2,2}A_n^2\exp(2i\omega  _0 n)+ \cc)
+a_{2,0}|A_n|^2\bigr) \non \\
&&+\ve^3(a_{3,3}A_n^3\exp(3i\omega _0 n)+\cc), \lb{solution5}
\eeqa
with a naive RG map for $A_n$
\beqa
A_{n+1}&=&A_n+i\ve^2(\delta+a_{3,1}|A_n|^2)A_n-i\ve^3\alpha_4\Aba_n^4
\lb{naive-rg5} \\
   &=&(1+\ve^2 {\cal L}_H+\cdots)A_n.
   \eeqa
A Hamiltonian generating the naive RG map (\ref{naive-rg5}) is given as
\beq
H(A,\Aba)=i \bigl(\delta |A|^2+\frac{a_{3,1}}{2}|A|^4-\ve
\frac{\alpha_4}{5}(A^5+\Aba^5)\bigr).
\lb{hamil-5}
\eeq
Although fixed points in the phase plane $(A_n,\Aba_n)$ are obtained
using the naive RG map
(\ref{naive-rg5}) or the Hamiltonian (\ref{hamil-5}), we derive them
through a symplectic
RG map.  In terms of the action-angle variables $(J_n,\theta _n)$, the
map (\ref{naive-rg5})
is easily made symplectic as
\beqa
J_{n+1}&=&J_n+\ve^2 F_5(J_{n+1},\theta_n) ,\non \\
\theta _{n+1}&=&\theta _n+\ve^2 G_5(J_{n+1},\theta_n) ,
\lb{sy-rg5}
\eeqa
where $F_5(J,\theta)=-2\ve\alpha_4 J^{5/2}\sin(5\theta)$ and
$G_5(J,\theta)=\delta+a_{3,1}J
-\ve \alpha_4J^{3/2}\cos(5\theta)$.
Fixed points of the map (\ref{sy-rg5}) are given by
\beq
\sin (5\theta)=0, \quad \delta+a_{3,1}J
-\ve \alpha_4J^{3/2}\cos(5\theta)=0,
\eeq
from which we have 10 fixed points
\beqa
J&\approx&-\frac{\delta}{a_{3,1}}+\ve \frac{\alpha_4
}{a_{3,1}}(-\frac{\delta}{a_{3,1}})^{3/2}, \qquad \theta
=\frac{2m\pi}{5} , \lb{fix-31}\\
J&\approx&-\frac{\delta}{a_{3,1}}-\ve \frac{\alpha_4
}{a_{3,1}}(-\frac{\delta}{a_{3,1}})^{3/2}, \qquad \theta
=\frac{(2m+1)\pi}{5} ,\lb{fix-32}
\eeqa
where $m=0,1,2,3,4$ and it is assumed that
\beq
-\frac{\delta}{a_{3,1}}>0. \lb{res-cond}
\eeq
The condition (\ref{res-cond}) simply states that the resonance
detuning $\delta$ should be
compensated by the nonlinear frequency shift.
A set of five fixed points (\ref{fix-31}) is found to be elliptic while
another set  (\ref{fix-31})
is hyperbolic. Therefore,  if the resonance detuning is set so that the
condition (\ref{res-cond})  is satisfied,
   the resonance structure for $2\pi/5$ resonance is composed of a chain
of five resonant islands.\\
   In the next section, the analysis in the present section is applied to
study the resonance structure of
   the H\'enon map.
%%%%%%%%%%%%%%%%%%%%%%%%%%%%%%%%%%%%%%%%%%%%%%%%%%%%%
\section{Application to H\'enon Map}
%%%%%%%%%%%%%%%%%%%%%%%%%%%%%%%%%%%%%%%%%%%%%%%%%%%%%%

   A map (\ref{basicmap}) with $a_m=0$ for $m>2$ seems to be rather
special
but has been extensively studied as the H\'enon
map\cite{henon}\cite{TD03}. Let us apply  results in the preceding
sections to the H\'enon map and compare them with numerical
calculations.%\\

In the case of  $2\pi/3$ resonance near the origin of the phase plane,
our analysis indicates that there are only three hyperbolic points of
period three and a chain of resonant island does not exist .
Numerical calculations of the H\'enon map in Fig.1 supports these results
of our analysis.
In the inner region of a triangle produced by connecting the three
hyperbolic periodic points,
the present analysis gives a good agreement   but is not able to
describe a chaotic sea in the outer region of the triangle.%\\

\begin{figure}[htb]
(a)\includegraphics[width=6cm,height=6cm]{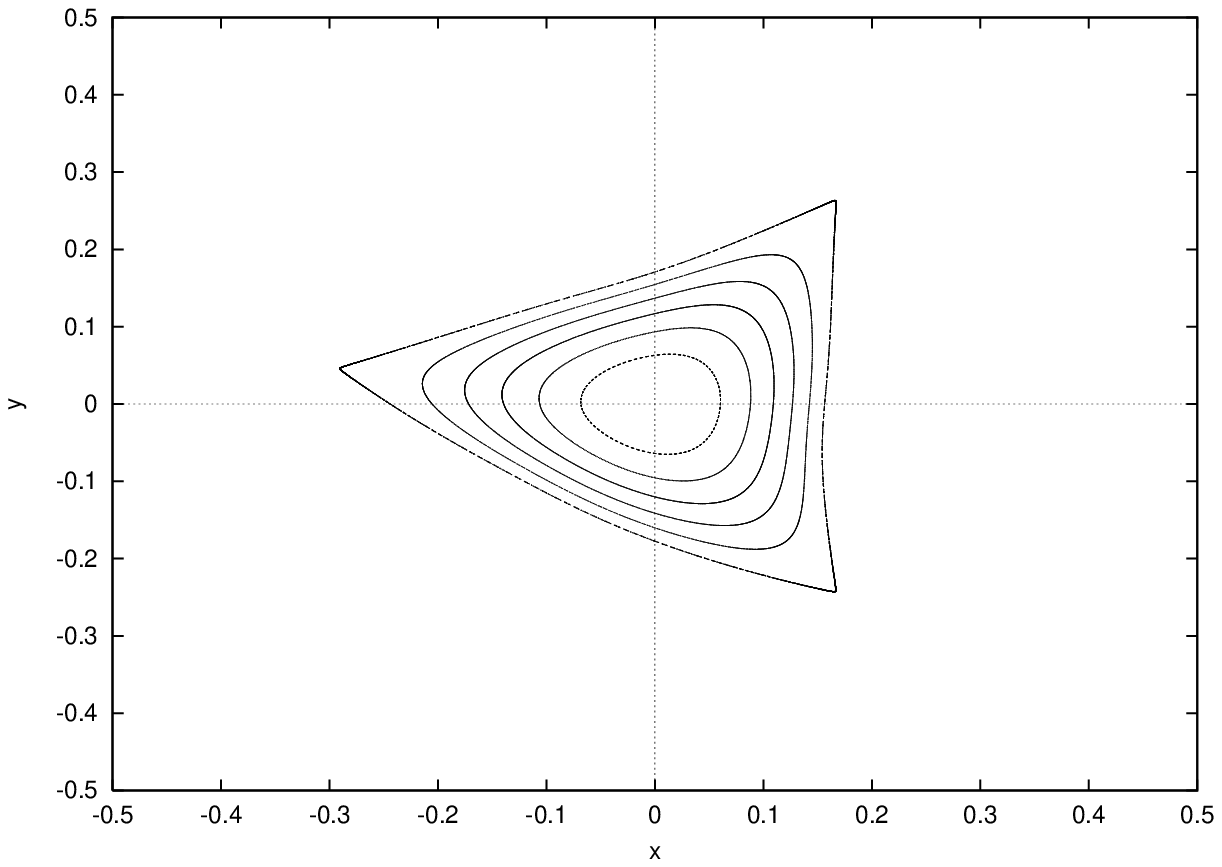}\quad
(b)\includegraphics[width=6cm,height=6cm]{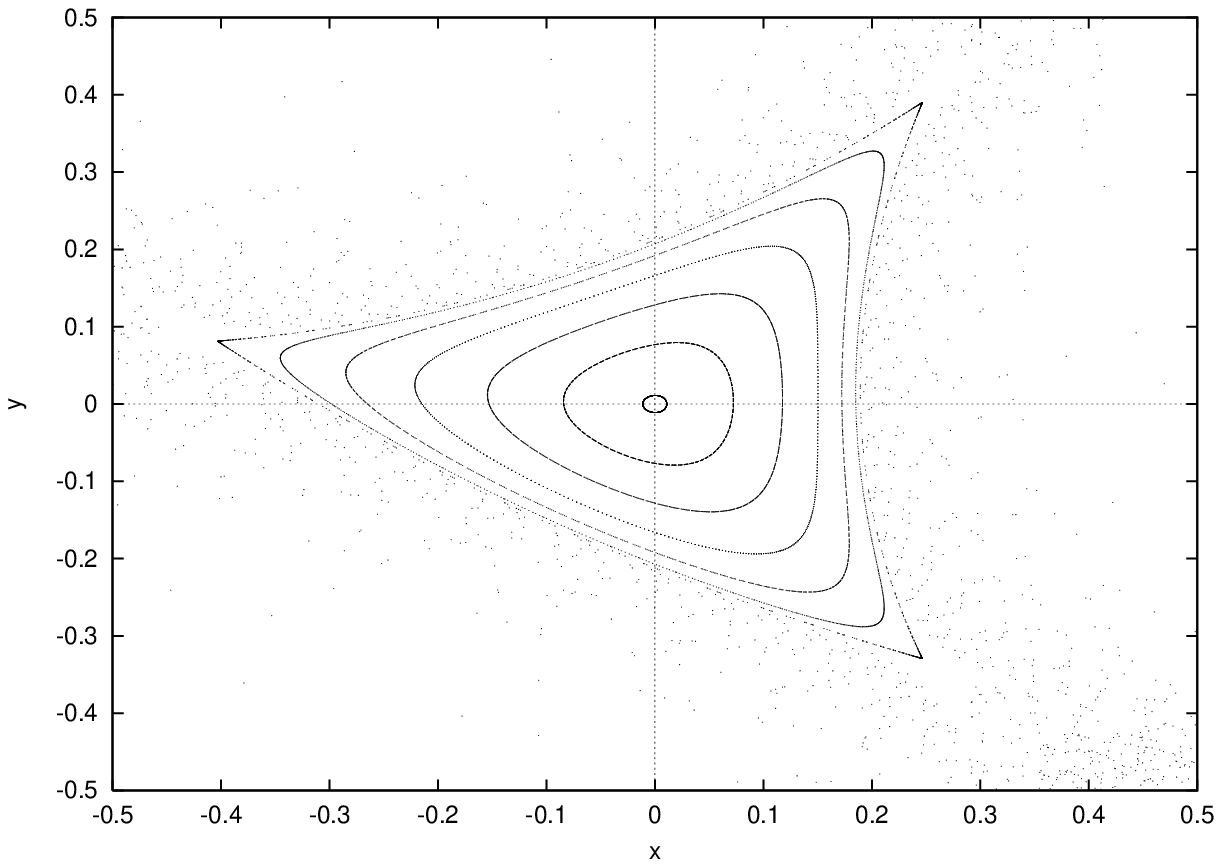}\\
Fig. 1. (a) phase portrait near the $2\pi/3$ resonance obtained from the 
RG map (\ref{sy-rg3}) for $\ve =0.1$ and $\delta =0.8$ \\
(b) phase portrait near the $2\pi/3$ resonance derived by
a numerical calculation of the H\'enon map for the same values of parameters.
\end{figure}

The case of  $2\pi/4$ resonance is exceptional in the sense that  the
former condition, which
guarantees existence of elliptic periodic points,
in (\ref{fix-cod1}) is not satisfied since $a_3=0$ for the H\'enon map.
Therefore, our analysis indicates that there are only four hyperbolic
fixed points near the origin of the phase plane. Numerical caluculation
confirms existence of the four hyperbolic fixed points but
there also exist four elliptic points far from the origin of the phase
space as shown in Fig.2.
Our perturbational analysis can not cover  a region far from the
origin. However,
if the third order correction to the RG map is taken into account, a
formal estimation shows that
four elliptic fixed points exist in a  far region, whose distance from
the origin is ${\cal O}(1/\ve)$.%\\

\begin{figure}[htb]
\begin{minipage}{7cm}
\vspace{4cm}
Fig. 2. phase portrait near the $2\pi/4$ resonance calculated from
the H\'enon map for $\ve =0.1$ and $\delta =0.8$ 
\end{minipage}
\begin{minipage}{6.5cm}
\begin{flushleft}
\includegraphics[width=6cm,height=6cm]{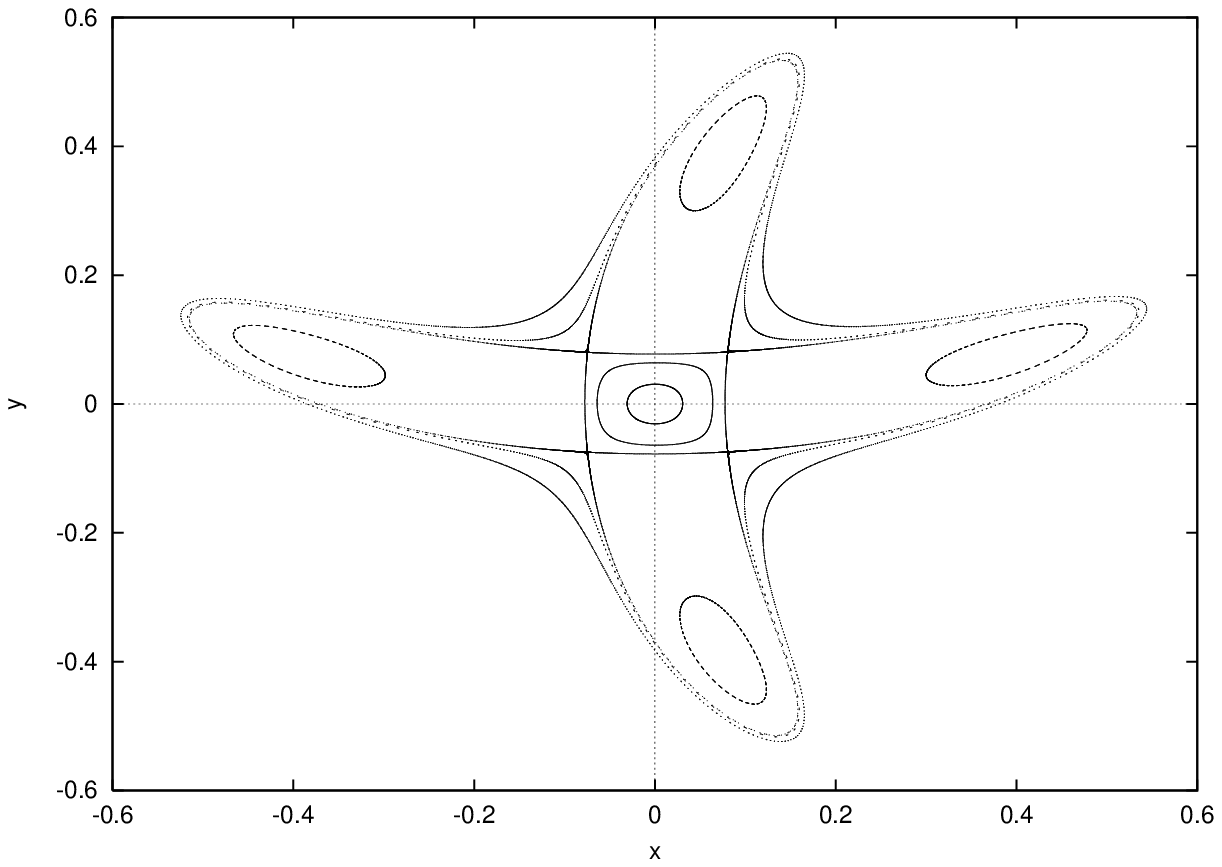}
\end{flushleft}
\end{minipage}
\end{figure}

In the case of  $2\pi/5$ resonance, we have a beautiful agreement of
the analysis and
numerical calculations of the H\'enon map as shown in Fig.3.
   While the phase portrait of the RG map has a symmetry of $2\pi/5$
rotation (Fig.3(a)),
   the phase structure in the H\'enon map does not have such symmetry (Fig.3(c)).
   This symmetry breaking is realized owing to the higher harmonic terms in the
solution (\ref{solution5}) as shown in Fig.3(b).

\begin{figure}[htb]
\begin{minipage}{7cm}
(a)\includegraphics[width=6cm,height=6cm]{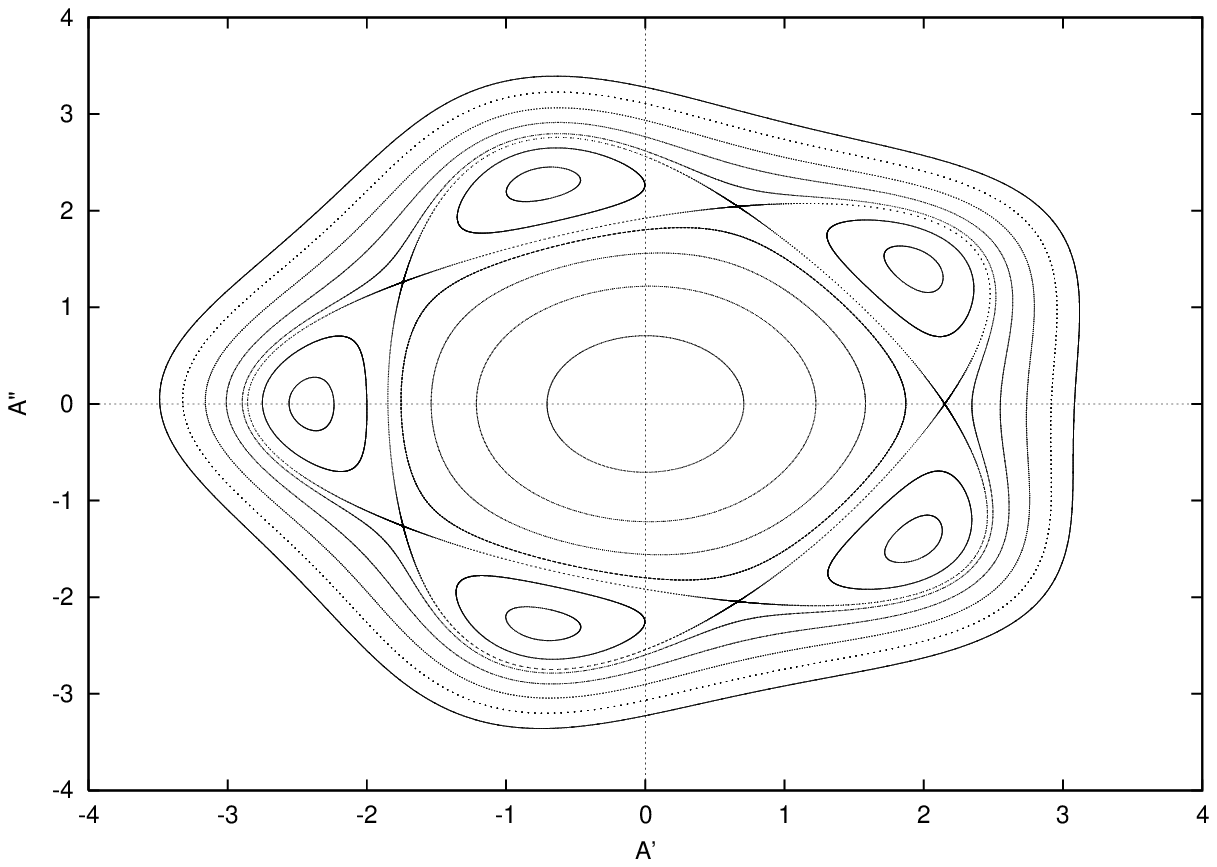}
\end{minipage}
\begin{minipage}{7cm}
(b)\includegraphics[width=6cm,height=6cm]{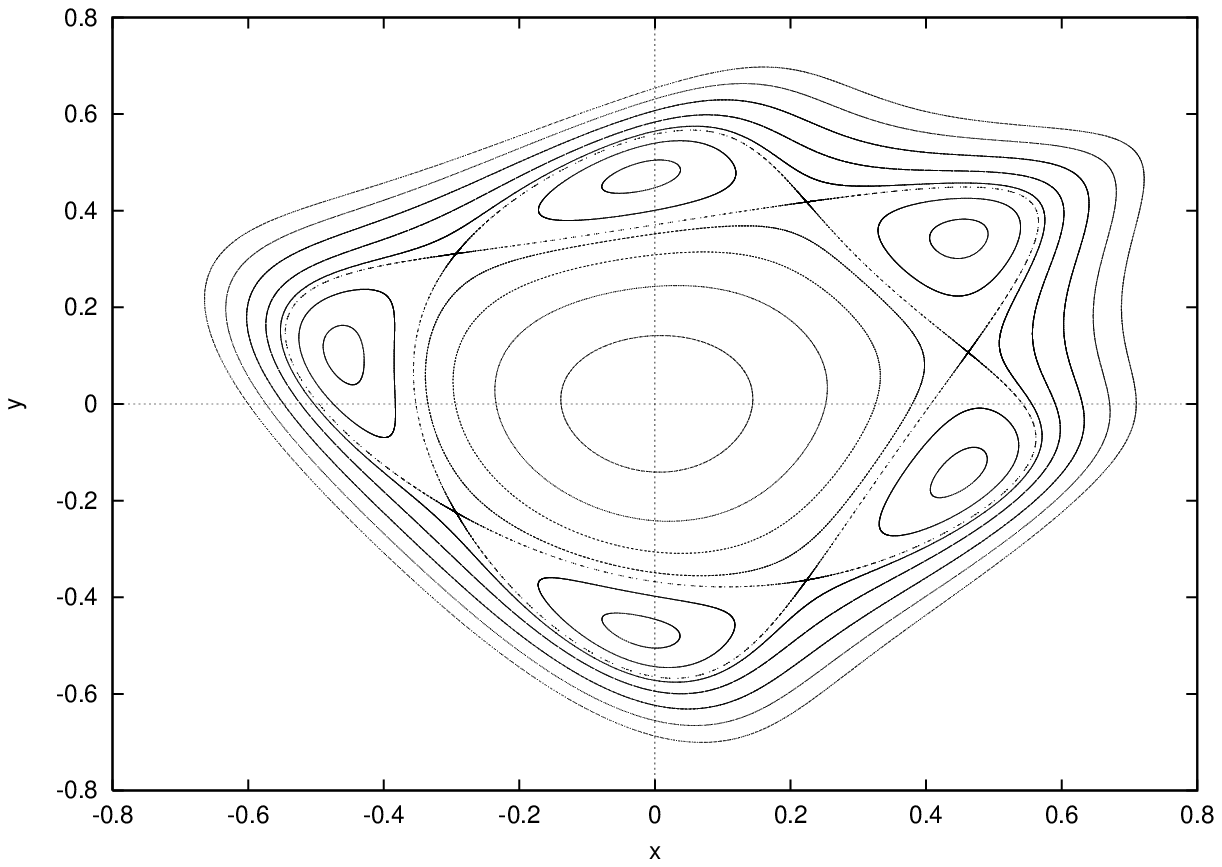}\\
\end{minipage}
\begin{minipage}{7cm}
Fig. 3. 
(a) phase portrait of the RG map (\ref{sy-rg5}) near the $2\pi/5$ resonance 
for $\ve =0.1$ and $\delta =0.8$, where $A^{'} =\sqrt{J}\cos(\theta)$ and
$A^{''} =\sqrt{J}\sin(\theta)$ \\
(b) phase portrait near the $2\pi/5$ resonance in the $(x, y)$ space derived
by the RG method\\
(c) phase portrait near the $2\pi/5$ resonance calculated from
the H\'enon map
\end{minipage}
\begin{minipage}{7cm}
(c)\includegraphics[width=6cm,height=6cm]{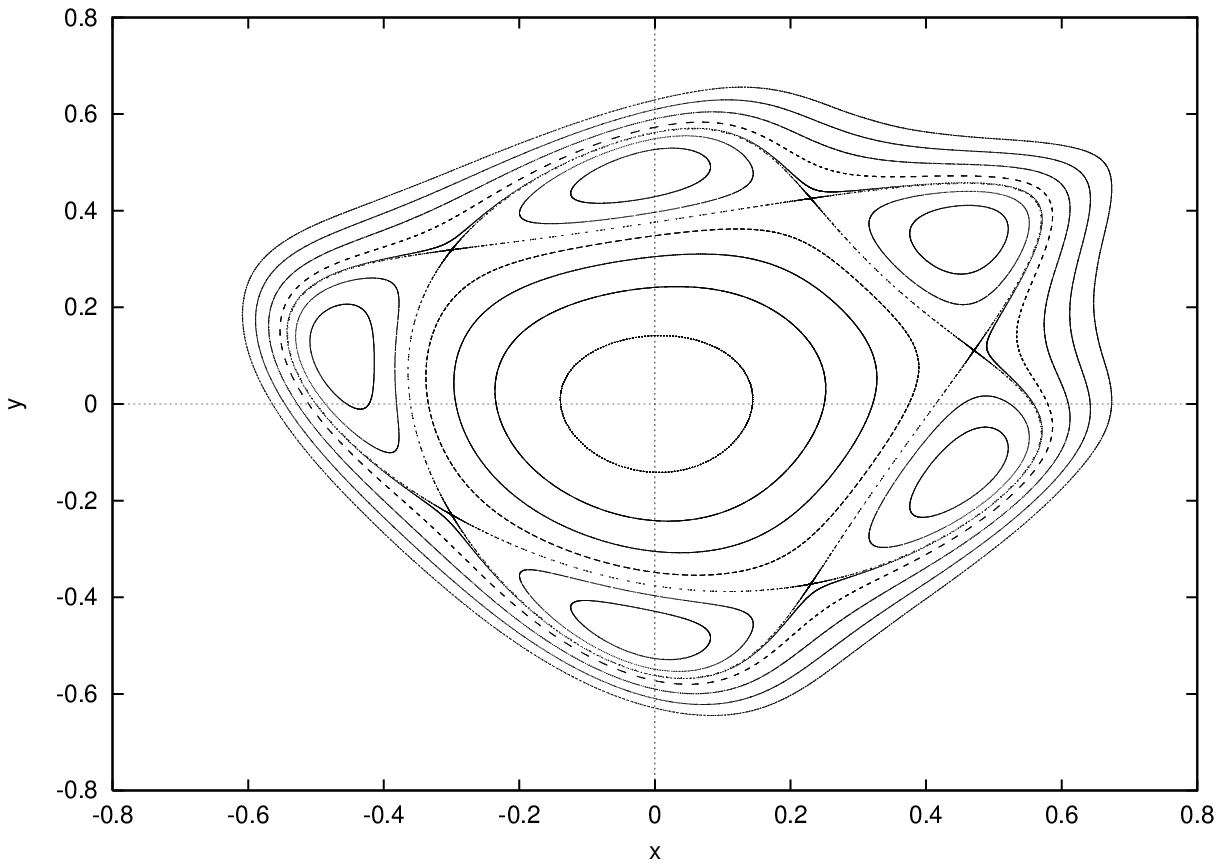}
\end{minipage}
\end{figure}

%%%%%%%%%%%%%%%%%%%%%%%%%%%%%%%%%%%%%%%%%%%%%%%%%%%%%
\section{Conclusion}
%%%%%%%%%%%%%%%%%%%%%%%%%%%%%%%%%%%%%%%%%%%%%%%%%%%
A symplecticity-preserving RG analysis is carried out  to study
resonance structures near an elliptic fixed point of a proto-type
symplectic map in two dimensions.  Symplectic RG maps are constructed
as  symplectic integrators to the underlying Hamilton flows.
Fixed points of the RG map correspond to periodic points of the
original map, which determine
   resonance structures such as a chain of resonant islands.  It should
be emphasized that we only need to analyze a distribution of fixed
points of the RG map in order to determine a topological structure of
the resonance region.
For  the resonant frequency $\omega _0 =2\pi/k  \quad (k=4,5)$,  the RG
map has $k$ hyperbolic
fixed points and $k$ elliptic fixed points around the origin with the
same angle spacing, while the case
$k=3$ is special because the RG map has only three hyperbolic fixed
points. Thus, we conclude
that  a chain of $k$ resonant islands constitutes the resonance
structures for $k=4,5$ but such a chain does not exists near the origin
for $k=3$.  Although we did not perform  analysis for
$k\ge 6$ in detail,  we are sure that the same conclusion as in the
case $k=4,5$ is valid in general.
In addition to the exisitence of fixed points, the RG map has
   a significant symmetry property that it is invariant with respect to
$\omega _0$ rotation around the origin, although the original
proto-type symplectic map does not posses such a
symmetry.  This rotational symmetry is broken due to the presence of
higher hamonics terms
and the symmetric RG map can reproduce  asymmetric resonance structures
in the original
phase space.

The above analytical theory for the proto-type symplectic map are
checked by numerical caluculations of the H\'enon map.  The result of
numerical calculations is shown to give a good agreement with the
theoretical result, especially for the case $k=5$.

%%%%%%%%%%%%%%%%%%%%%%%%%%
\section*{Acknowledgement}
%%%%%%%%%%%%%%%%%%%%%%%%%%
One of the authors (S.G.) has been supported by
a JSPS Fellowship for Young Scientists.
Another author (K.N.) has been, in part, supported by a Grant-in-Aid for
Scientific Research
(C) 13640402 from the Japan Society for the Promotion of Science.
%%%%%%%%%%%%%%%%%%%%%%%%%%%%%%%%%%%%%%%%%%%%%%%%%%%%%%%%%%%%%%%%

%%%%%%%%%%%%%%%%%%%%%%%%%%%

%%%%%%%%%%%%%%%%%%%%%%

%%%%%%%%%%%%%%

\begin{thebibliography}{99}
%%%%%%%%%%%%%%%%%%%%%%%%%%%

\bibitem{CGO96}L. Y. Chen, N. Goldenfeld and Y. Oono,
Phys. Rev. {\bf E54} (1996), 376.
% ``Dynamical reduction of discrete systems based on the
%      renormalization-group method''
%
\bibitem{GMN99}S. Goto, Y. Masutomi, and K. Nozaki,
Prog. Theor. Phys. {\bf 102} (1999), 471.
% ``Lie-Group Approach to Perturbative Renormalization Group Method''
%
%8
\bibitem{GN01Prog} S. Goto  and K. Nozaki,
Prog. Theor. Phys. {\bf 105} (2001), 99.
% ``Asymptotic Expansions of Unstable and Stable Manifolds in
%      Time-Discrete Systems''
%
%9
\bibitem{GN01JPSJ} S. Goto  and K. Nozaki,
J. Phys. Soc. Jpn. {\bf 70} (2001), 49.
% ``Regularized Renormalization Group Reduction of Symplectic Maps''
%

%
\bibitem{KM98}T. Kunihiro, J. Matsukidaira, Phys. Rev. {\bf E57}
(1998), 4817.
% ``Dynamical reduction of discrete systems based on the
%      renormalization-group method''
%
%\bibitem{Tze01} S.I. Tzenov,
%{\it preprint,}~{\tt  http://jp.arXiv.org/abs/physics/0106101} (2001)

%
% ``Renormalization Group Approach to the Beam-Beam
%    Interaction in Circular Colliders''
%
\bibitem{GNY02}S. Goto, K. Nozaki, and H. Yamada,
Prog. Theor. Phys. {\bf 107} (2002), 637.
% ``Random Wandering around Homoclinic-Like Manifolds
%                               in a Symplectic Map Chain''
\bibitem{henon} M. H\'enon, Quart. appl. Math. {\bf 27} (1969), 291. 


\bibitem{TD03}S. I. Tzenov and R.C. Davidson,
New Journal of Physics {\bf 5} (2003), 67.
%
%\bibitem{Tze02}S. I. Tzenov,
%New Journal of Physics {\bf 4} (2002), 6.
%

\bibitem{rgtrans}
It should be noted that a continuous version of the simple RG 
transformation used here was first introduced in \cite{GMN99} and is 
different from the original RG transformation
by Chen, Goldenfeld and Oono  \cite{CGO96}.  Its discrete version was 
first used in \cite{GN01JPSJ}


\bibitem{GN03} S.Goto and K.Nozaki, 
http://xxx.lanl.gov/abs/nlin.CD/0309021.

%%%%%%%%%%%%%%%%%%%%%%
\end{thebibliography}
\end{document}